\documentclass[prb,twocolumn,floats]{revtex4}
\usepackage{epsfig}
\usepackage{amsmath}
\usepackage{graphicx}
\usepackage{soul}
\usepackage{setspace}
\usepackage{color}

\begin{document}
\title{\bf Quantum Oscillation and Landau-Zener Transition in {Untilted} Nodal Line Semimetals under a Time-Periodic Magnetic Field}
\author{Satyaki Kar}
\affiliation{{A.K.P.C. Mahavidyalaya, Bengai, Hooghly-712611, West Bengal, India}}
\begin{abstract}
  Nodal line semimetals (NLSM) exhibit interesting quantum oscillation characteristics when acted upon by a strong magnetic field. We study the combined effect of strong direct (dc) and alternating (ac) magnetic field, perpendicular to the nodal plane {in an untilted NLSM in order to} probe the behavior of the low lying Landau level (LL) states that can periodically
  become gapless for suitably chosen field parameters. The oscillatory field variation, as opposed to a steady one, has interesting impact on the quantum oscillation phenomena with the Landau tubes crossing the Fermi surface extremally two times per cycle.
Furthermore, the low energy modes can witness Landau-Zener like transitions between valence and conduction band providing further routes to conduction. We discuss such transition phenomena following the framework of adiabatic-impulse approximation for slow quenches. 
Next we also investigate the effect of oscillating magnetic field acting parallel to the nodal loop where topologically nontrivial magnetic oscillations at low energies can be witnessed. Therefore, with proper parameters chosen, one can engineer topological transitions to occur periodically in such systems as the oscillating field is swept through its cycles.
\end{abstract}

\maketitle                              

\section{Introduction}
In the recent surge of studies involving topological condensed matter systems, a nodal-line semimetal (NLSM) has become a familiar name these days.
{Usually in a topological semimetal, band crossings occur at discrete points or along lines/loops within a Brillouin zone. In the former scenerio, one finds non-degenerate (doubly degenerate) band-crossings of Weyl semimetals (Dirac semimetals) while the later scenerio results in an NLSM\cite{ashvin}.}
A NLSM has accidental band touching nodes of codimension 2 with conduction and valence band crossing in a 3D Brillouin zone at specific symmetry protected line nodes\cite{burkov} and is characterized by nonzero topological invariants\cite{rev1}. Not only are such systems proposed theoretically\cite{fang,grphne,IScrystal}, they are realized experimentally{\cite{cu3pdn,pbtase2,titase2,zrsis,zrsise,photonic}} as well.
  The stability of the nodal loops/rings in a NLSM usually requires absence of spin-orbit couplings (SOC), though there are also propositions for stable line nodes in presence of SOC\cite{fang,rev1,ashvin}. {In general, the local Weyl cones on the nodal ring can have a tilted orientation, based on which one can distinguish between a type-I (small tilt) and a type-II (large tilt) NLSM material\cite{njp}. Out of those, only an untilted NLSM} contains equi-energy line nodes\cite{njp}. Such gapless spectrum gets split into quantized Landau levels (LL) under the application of a strong magnetic field\cite{lim1,molina,rev2,lim2,B@x,njp}. Conducting surface states, however, remain present turning the system into a topological insulator (TI)\cite{molina}. These quantized spectra show interesting variation for different orientations of the magnetic field about the nodal plane\cite{lim2,B@x}.
A Fermi electron's accumulation of a Berry phase around a closed loop becomes topological or trivial depending on the direction of the field{\cite{berry}}.
These systems exhibit quantum oscillations (QO) as a steady variation of the field strength results in periodic variation of the density of states{\cite{lim2,balents,cortijo}} and henceforth susceptibility, resistivity, magnetoresistance etc. - their phases being also dependent on the Berry phases corresponding to the electronic cyclotron motion\cite{lim2,cortijo,li}.

{Now notice that a periodic driving can produce many nontrivilities like stuckelberg interferences, dynamic freezing or Floquet engineering to a system\cite{lz,kar1,kar2,floquet}. A NLSM produces a Weyl semimetallic Floquet spectrum upon irradiation via circularly polarized light and thereby contribute to photovoltaic anomalous Hall effect\cite{photo2,photo}. Similarly, a driven 3D magnonic Dirac nodal-line can also produce Weyl magnons\cite{photo-m}. In the present context,} we find that by adding an alternating (ac) field, just as an envelope to the strong direct (dc) magnetic field already present in a NLSM system, we can easily tune the QO phenomena.
Like a Landau tube pops out of the Fermi surface by a steady variation of the magnetic field, an oscillating field variation causes oscillating changes in Landau tube dimensions and allows it to possess the  extremal cross-sections two times, if not none, per cycle of the field sweep. This affects the temporal periodicity of the quantum oscillations.  Instead of searching for the occasional gapless states, a periodic field variation can also let the system pass through gapless phases periodically for proper choice of the field parameters.
Even in presence of a gap, transition of electrons from valence band to conduction band becomes possible due to multiple passage through the avoided crossing points\cite{lz} of the spectrum. It disrupts the insulating nature of the low lying gapped states. Interesting exciting patterns can be observed due to Stuckelberg interference from all possible paths of transitions in the two-level system. One can understand the slow driving and fast driving scenario, in this regard, under the framework of adiabatic-impulse approximation\cite{lz,kar1,kar2} and rotating wave approximation\cite{lz,kar2}/Floquet theory\cite{floquet} respectively.

{To name a few NLSM candidates, one can mention the inversion symmetric compound $Cu_3PdN$\cite{cu3pdn}, noncentrosymmetric $(Pb,Ti)TaSe_2$\cite{pbtase2,titase2}, nonsymmorphic $ZrSiS$\cite{zrsis}, $ZrSi(Se,T)$\cite{zrsise}, artificial 2D nonsymmorphic photonic crystal lattices\cite{photonic} and many more.
Applying an oscillating magnetic field on such NLSM systems can be established either directly or with the help of irradiation\cite{dini,photo2,photo}.} The resulting novel transport characteristics can thus be easily examined.
This paper provides an analytical and numerical study of such problems where we consider the oscillating magnetic field not only perpendicular to the nodal plane but also parallel to it, as the later can give rise to topological quantum oscillations at low energies\cite{lim2,cortijo}.
{The paper is organized as follows. In section II, we provide the formulation and spectral analysis of the problem for magnetic field perpendicular to nodal rings. Section III and IV discuss the corresponding quantum oscillation characteristics and intra/inter band transitions respectively. Then in section V, we briefly give the formulation for the case when the field is parallel to the nodal plane and discuss the corresponding QO phenomena. Finally we conclude our findings elaborating on possible future works in section VI.}

\section{Formulation and Spectra}
We consider a simple continuum model NLSM Hamiltonian with time-reversal and space-inversion symmetries\cite{photo} to be given as
\begin{eqnarray}
H_0=(\frac{p_\perp^2}{2m}-m_0)\sigma_z+vp_z\sigma_y
\label{eq1}
\end{eqnarray}
where $p_\perp^2=p_x^2+p_y^2$.
The $\sigma$ matrices are the Pauli matrices describing orbital degrees of freedom {(called pseudospins)} while we omit the spin degrees of freedom for this model possesses no SOC. It represents a nodal circle of radius $\sqrt{2mm_0}$ in the $p_z=0$ plane. 
On applying a magnetic field ${\bf B}=(0,0,B)$ perpendicular to the nodal loop, the Hamiltonian gets modified via Peierls substitution{\cite{allen,goswami,dini,photo}} ${\bf p}\rightarrow{\bf p-eA}$ {(see Appendix-A)}. With vector potential ${\bf A}$ in asymmetric Landau gauge being given as ${\bf A}=(-By,0,0)$, the Hamiltonian becomes
\begin{eqnarray}
  H&=&H_0+\frac{e}{m}(-p_xBy+\frac{eB^2y^2}{2})\sigma_z.
  \label{eqB}
\end{eqnarray}
In the basis of Landau states this can be written as
\begin{eqnarray}
H(B,p_z)=[(n+\frac{1}{2})\frac{e B\hbar}{m}-m_0]\sigma_z+vp_z\sigma_y
\label{eqLL}
\end{eqnarray}
with $\epsilon_{n,p_z}=\pm\sqrt{[(n+\frac{1}{2})\frac{e B\hbar}{m}-m_0]^2+v^2p_z^2}$ being the dispersions. {So there are two dispersion branches  (say, $(+)$ and $(-)$) corresponding to each Landau level $n$. Here the Landau level states are free particle wavefunctions multiplied by eigenstates for displaced harmonic oscillators\cite{molina,fazekas}} {(see Appendix-A)}.

Now since $p_z$ is not affected by ${\bf B}$, the 3D problem decomposes into a family of 2D ones parameterized by $p_z${\cite{lim2}}.
And for each $p_z$, we find discrete Landau level states with huge degeneracy that is also proportional to the magnetic field strength B\cite{ashcroft,fazekas}.

\begin{figure}
\centering
\includegraphics[width=\linewidth,height=2.5 in]{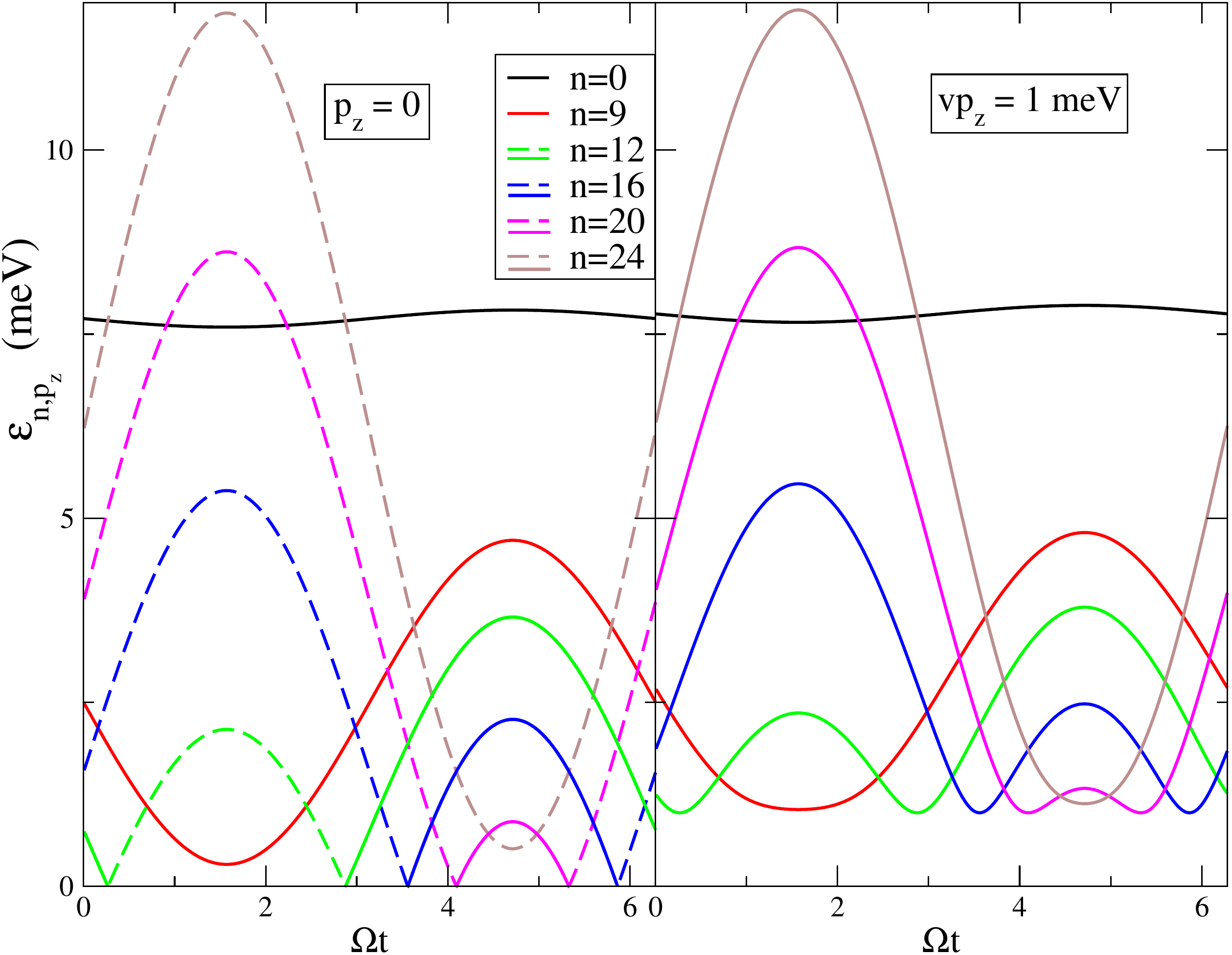}
\caption{(Color online) Plot of the energy dispersions for low energy Landau states at different $p_z$ values for $B_0=5 T,~B_1=2 T$ and $m_0=8~meV$. {The solid and dashed lines are used to distinguish between $(+)$ and $(-)$ branches of each LL.}}
\label{fig1}
\end{figure}

Few low lying spectra including the $n=0$ LL are shown in Fig.\ref{fig1} as functions of time within single cycle of the field $B$. The solid and dashed lines distinguish between two branches of the spectra $\epsilon_{n,p_z}$.
Notice that, for a particular $p_z$, minimum energy does not necessarily correspond to $n=0$. Rather, it corresponds to an integer that is closest to the expression $\frac{mm_0}{eB\hbar}-\frac{1}{2}$.

Out of the spectra shown in Fig.\ref{fig1}, only $n=0$ and $9$ LL states are gapped at $p_z=0$.
A zero energy mode with $\epsilon_{n,p_z}=0$ appears only when $p_z=0$ and $(n+\frac{1}{2})\frac{e B\hbar}{m}=m_0$, a condition which is generally not met, for $n$ is an integer. Thus bulk states remain gapped.
But a combination of ac and dc field, such as $B(t)=B_0+B_1sin(\Omega t)$, can produce gapless bulk Landau states two times per cycle whenever $\frac{mm_0}{e B(t)\hbar}-1/2$ becomes an integer.
The band touching (possible for $p_z=0$ alone) again takes the form of nodal circles. {We should mention here that we ignore the electric field produced by the time dependent magnetic field and accordingly contain ourselves to small $\Omega$ values alone} {(see Appendix-A)}. 

In our formulation we don't consider the ac field alone as the Zeeman effect becomes non-negligible whenever $B(t)$ becomes comparable to it during its cycle.
Now even if there is no band touching ($e.g$, for $p_z\ne0$) due to this ac field sweeping, Landau-Zener (LZ) transitions between the low energy modes can cause periodic avenues for charge transport from valence band (VB) to conduction band (CB).

\section{Quantum Oscillations}
{In a 3D system, quantized Landau tubes of electron motion are formed in presence of a magnetic field ${\bf B}$.} The semi-classical equations of motion for Bloch electrons in the crystal produce electronic cyclotron orbits with angular frequency $\omega_c=eB/m^*$, $m^*$ being the effective mass of electron in the periodic lattice.
Onsager's theory shows that such orbits in the momentum space ($i.e.,~{\bf p}$-space) are quantized with area $A_n=(n+\gamma)2\pi\hbar eB$ (and thus $A_{n+1}-A_n=2\pi\hbar eB$) for LL dispersions close to Fermi energy\cite{ashcroft}. This is because the energy difference $E_{n+1}(B,p_z)-E_n(B,p_z)=\hbar eB/m^*$ ($z$ being the magnetic field direction) under same conditions. One finds that $m^*=\frac{1}{2\pi}(\partial A_n/\partial E)$ and $\gamma$ is a phase parameter related to Berry phase $\gamma_B$ for the orbital motion as {$\gamma=\frac{1}{2}\pm\frac{\gamma_B}{2\pi}$\cite{cortijo,phase}.}

As $B$ is varied, cyclotron orbits grow or shrink in sizes and for a steady increase of B, more and more Landau tubes pop out of the Fermi surface gradually. {Whenever such a popping out takes place, one gets $E_n(B,p_z)=E_F$, the Fermi energy. However, only for few $B$ and $p_z$ values, the cyclotron orbit cross-section becomes extremal, $i.e.,$ $\partial A_n/\partial p_z=0$. This also indicates accumulation of large number of states or high DOS.}
Furthermore, as $B$ is steadily increased/decreased, such jump in DOS occurs periodically. This gives quantum oscillation in the system because a constant change in $B^{-1}:~\Delta(1/B)=\frac{2\pi\hbar e}{ A_{ex}}$ causes the DOS peak to appear again as Landau tubes of successive orders (from $n$ to $n\pm 1$, say) cross the Fermi surface with extremal area $A_{ex}$ (which is a function of $E_F$ and not of $n$).
\begin{figure}
\centering
\includegraphics[width=\linewidth,height=2. in]{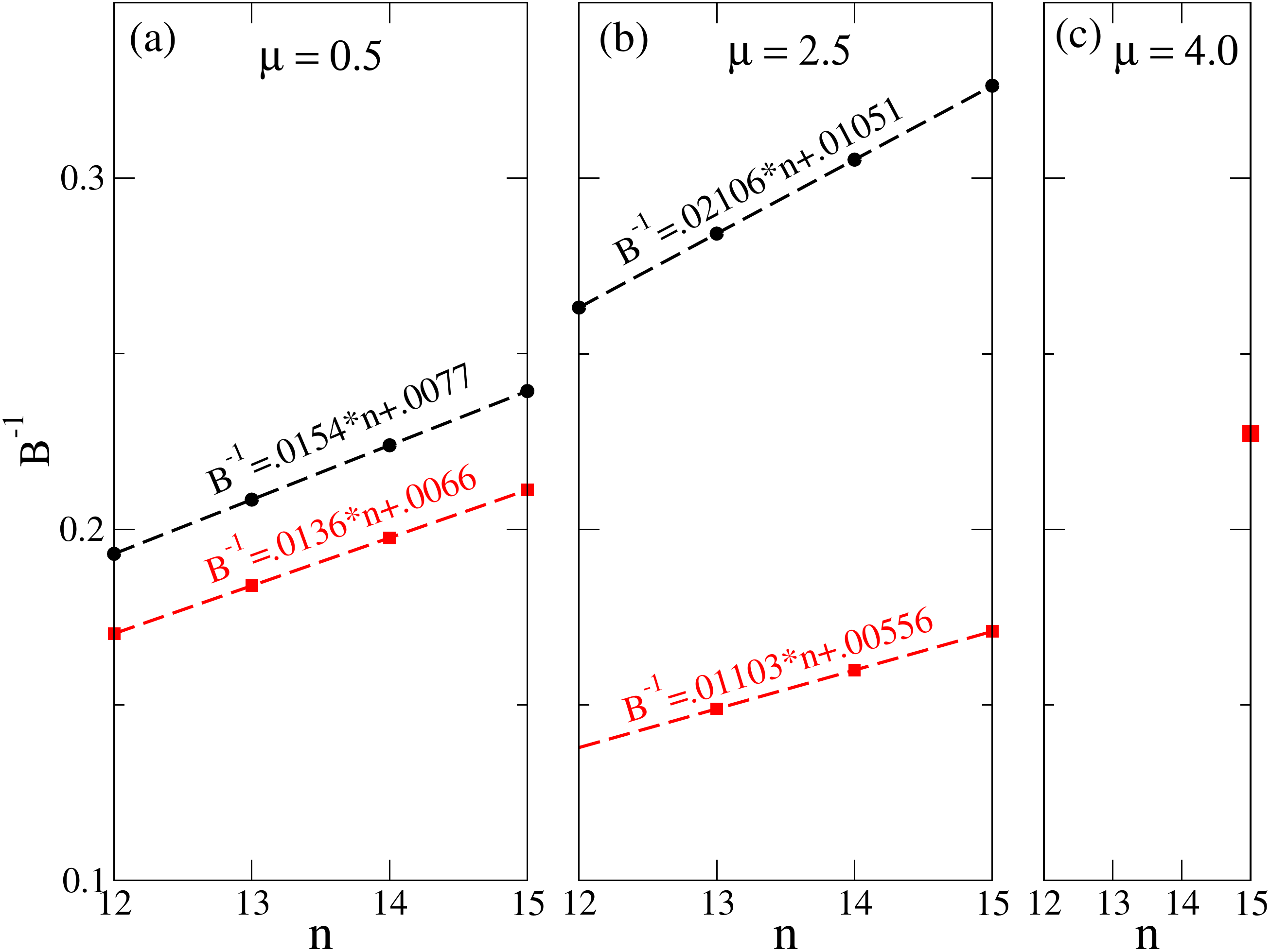}
\caption{(Color online) Landau fan diagrams at $p_z=0$ for parameters $B_0=5~T,~B_1=2~T$ and $m_0=8~meV$. It shows $B^{-1}$ vs. $n$ plots for $n\rightarrow 12-15$ and $\mu=0.5,~2.5~\&~4.0~meV$. {The black (circles) and red (squares) points denote extremal orbit coordinates from $(+)$ and $(-)$ branches of the spectrum respectively.} }
\label{lfan}
\end{figure}

Let us now consider the case of time periodic field $B(t)=B_0+B_1sin(\Omega t)$
acting on our NLSM system. Particularly if $\frac{2B_1}{B_0^2-B_1^2}\geq \frac{2\pi e}{\hbar A_{ex}}$, quantum oscillation of the previous kind can be observed.
Extremal cross-section $A_{ex}$ is obtained at $p_z=0$ for there one finds $\partial A_n/\partial p_z=0$. Thus we stick to the $p_z=0$ plane as far as the discussion on the quantum oscillation is concerned.

{For ${\bf B}\perp$ nodal plane, $n$th LL spectrum becomes gapless at a singular value of $B$, say $B_c(n)$, where both its dispersion branches cross each other. Such gaplessness can be seen in our system only if $B_0-B_1<B_c(n)<B_0+B_1$. Otherwise the LL state will appear as a gapped state.
So for $\mu=0$, a gapless LL state gives two identical extremal cyclotron orbits corresponding to two branches of its dispersions whereas a gapped state does not show any QO behavior.}

{At this point we should mention that the Fermi level or chemical potential $\mu$ also oscillates with a field variation which is ignored in many the theoretical calculations\cite{molina,lim2,cortijo} including ours. This is because such changes as well as their effect are usually small compared to the effect of cyclotron orbit broadening or shrinking due to $B$ variation (see Lifshitz-Kosevich formulation in \onlinecite{mu-constant}). However, there are some exceptions as well\cite{mu-oscillation}.
Though in the present paper, we maintain the constancy of $\mu$ with field variation, we plan to turn to this issue in a future communication.}
  
{Let us now consider the case for finite $\mu>0$. For a gapless LL state, the $(+)$ and $(-)$ branches cross the $\epsilon_n(B,p_z=0)=\mu$ line at two different values of $B$, say $B_\mu^{(+)}$ and $B_\mu^{(-)}$. So one gets different extremal orbits corresponding to different branches of dispersions. These two sets of extremal orbits will have different areas: one smaller and one larger than the extremal area obtained at $\mu=0$. Contrarily for a gapped state, the $(-)$ branch never touch the Fermi surface. The $(+)$ branch, however, can cross the Fermi level if $\mu$ falls within its bandwidth. In that case, we get a single extremal orbit corresponding to $B_\mu^{(+)}$. Saying in other words, there are two critical chemical potentials, say $\mu_n^{(1)}~\&~\mu_n^{(2)}$ for the $n$th LL such that it can show QO phenomena whenever $\mu_{n}^{(1)}<\mu<\mu_{n}^{(2)}$.}
\begin{figure}
\centering
\includegraphics[width=\linewidth,height=3. in]{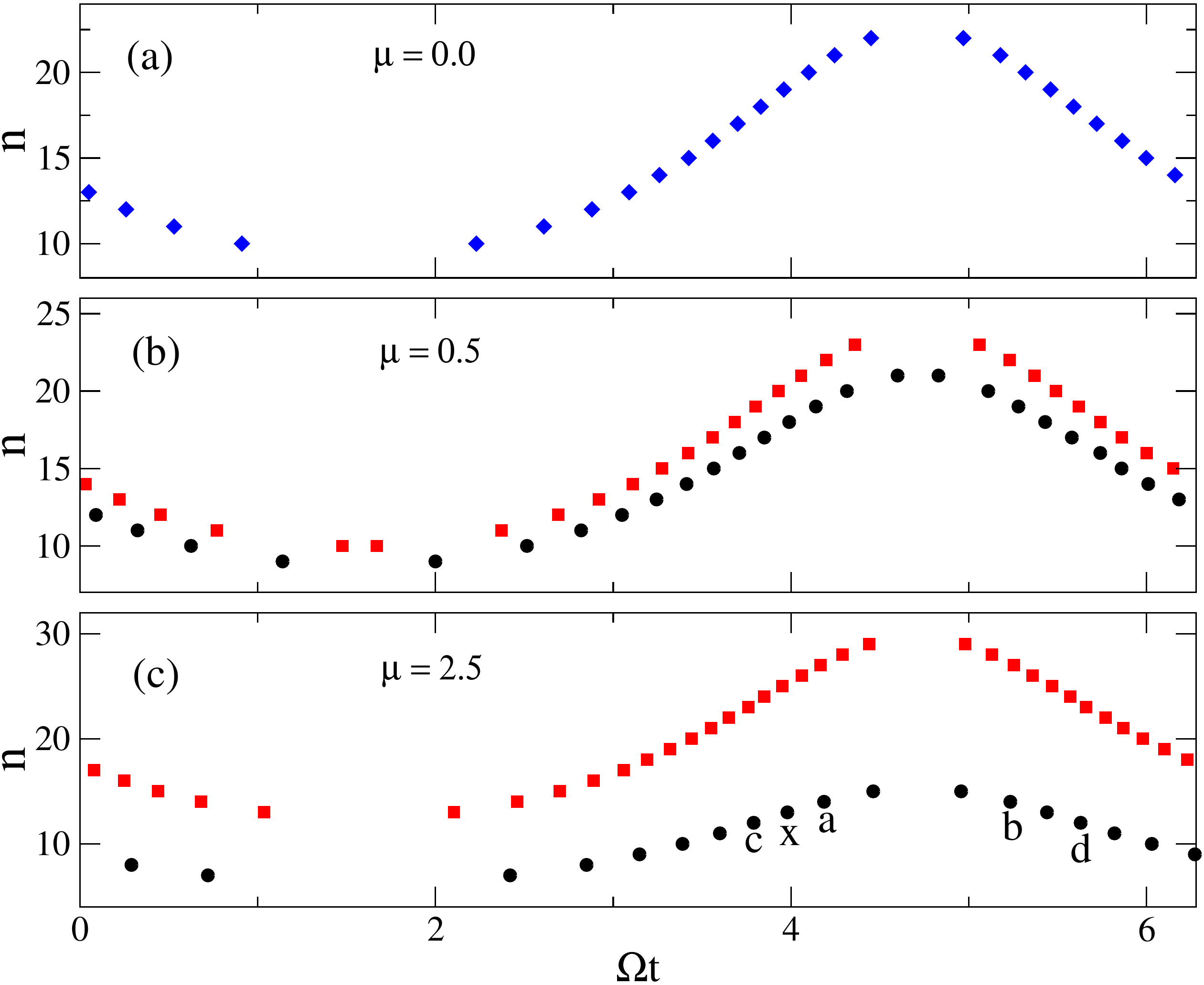}
\caption{(Color online) $n$ vs. $\Omega t$ plots for $\mu=0,~0.5~\&~2.5~meV$ and field parameters same as used in Fig.\ref{lfan}. \\{(b),(c): The black (circles) and red (squares) points denote extremal orbit coordinates from $(+)$ and $(-)$ branches of the spectrum respectively. (a): The blue (diamonds) points refer to the $\mu=0$ case where two sets of points corresponding to $(+)$ and $(-)$ branches merge to each other.} }
\label{n-t}
\end{figure}

{In a typical QO phenomena, repeated appearance of extremal orbits is registered for a variation of field given as $\Delta(1/B)=\frac{2\pi e}{\hbar A_{ex}}$. Thus one obtains linear plots in $B^{-1}$ vs. $n$ Landau fan diagrams where each point corresponds to an extremal orbit. 
Now notice that in the periodic protocol like that we use, an individual $B$ value appears twice per single cycle/sweep of the field. So when viewed over time, we get a pair of extremal points at a LL dispersion branch (for $\mu$ within its bandwidth) in every cycle of the field sweep.}



All these conjectures can be understood from the Landau fan diagrams in Fig.\ref{lfan}. There the $B^{-1}$ versus $n$ plots describe uniform periodicity in quantum oscillations with two sets of points ($i.e.,$ circles and squares) referring to extremal orbits coming from two different branches of LL spectrum. {Notice that there are two separate QO frequencies (which equals the slope for these lines) with a larger $A_{ex}$ value for the $(-)$ branch and the vice versa. Moreover, the $n$-intercepts $\sim~-0.5$ indicate the compatibility of the fan diagrams with the Onsager's relation with $\gamma=\frac{1}{2}$ or $\gamma_B=0$.} So for $B\perp$ nodal-plane, topologically trivial oscillations are obtained where the cyclotron orbits never cut through the nodal loop\cite{lim2}. Notice that, at a given $B$ value, the number of extremal orbits in a LL state depends on the value of $\mu$.
{Now these linear trends of fan diagrams show non-uniformities in QO periodicity when viewed as a function of time. This has been demonstrated in Fig.\ref{n-t}.}
The time periodicity in $B(t)$ in combination with periodicity in DOS as a function of $B^{-1}$ results in interesting modification to the quantum oscillation behavior.
{First of all realize that for $\mu=0$, each point in a fan diagram indicates two pairs of identical extremal orbits each corresponding to the $(+)$ or $(-)$ gapless LL branches. However, each pairs of orbits corresponding to one branch appear at different times as can be seen in Fig.\ref{n-t}(a). Thus each point in Fig.\ref{n-t}(a) correspond to two extremal orbits from two branches.
But for $\mu>0$, non-identical extremal orbits are obtained corresponding to two branches of a LL as they appear at different $B$ values.
Therefore each point of Fig.\ref{n-t}(b)-(c) diagrams correspond to a single extremal point.
Starting from such an extremal point in a LL, say $n$, now
there will be 4 time-steps $\Delta t(n)$ within a cycle (instead of 2, for a non-oscillating $B$ variation), that will correspond to $\Delta(1/B)=\frac{2\pi\hbar e}{ A_{ex}}$ or $\Delta n=\pm 1$ in the same LL dispersion branch. However, this requires $\mu$ to fall within the bandwidth of the spectra of the neighboring Landau levels. For example, Fig.\ref{n-t}(c) highlights a typical extremal point $x$ at $n=13$ and its 4 neighboring points $a,b,c$ and $d$ corresponding to $n=13\pm1$.}





\section{Intra/Inter band transitions}
{In a semiclassical model, electrons can only move within a band without any transition to other bands\cite{ashcroft}. However for a zero or small energy gap as that we encounter in this problem, inter/intra-band transitions between valence and conduction bands corresponding to same (intra) or different (inter) Landau levels need to be considered}.
Let's consider the Hamiltonian $H(B,p_z)$.
We can write it as
\begin{eqnarray}
  H=[C_1+C_2sin(\Omega t)]\sigma_z+C_3\sigma_y
  \label{eq2}
\end{eqnarray}
where $C_1=(n+\frac{1}{2})\frac{e B_0}{m}-m_0$, $C_2=(n+\frac{1}{2})\frac{e B_1}{m}$ and $C_3=vp_z$ and $\sigma$'s denote the orbital space (obtained from VB and CB). Let us stick to the low energy part of the spectrum and consider only those $n$ for which the gap between the bands, $i.e.,~2|\epsilon_{n,p_z}|$ is small.
The intraband transitions in such two level systems {(TLS)} can be obtained numerically. However in order to understand the behavior, an analytical framework is always desirable. Hence we describe the problem in terms of Adiabatic-Impulse approximation\cite{lz,kar1,kar2}, which works well in small $\Omega$ limit ($i.e.,~C_2\Omega<<\epsilon_{n,p_z}^2$).

In the two level system {given by Eq.\ref{eq2}}, each cycle/sweep corresponding to $C_2>|C_1|$ gives a pair of avoided crossing points (ACP)\cite{lz} at times $t_1$ and $t_2$ given by
\begin{align}
  \Omega t_1&=sin^{-1}(-\frac{C_1}{C_2})~~~\&~~~\Omega t_2=\pi-\Omega t_1.\nonumber
\end{align}
However there is only one ACP given by $\Omega t_1=-sgn(C_1)\frac{\pi}{2}$ for $C_2\le|C_1|$.
Close to ACPs, the problem becomes a Landau Zener (LZ) problem (for $C_2>|C_1|$) which is given by a linear time dependent Hamiltonian $H=vt\sigma_z+\Delta\sigma_x$ with intraband transition probability $P_{LZ}=exp[-\pi \Delta^2/v]$\cite{lz}. In the present case, this takes the form $P_{LZ}=exp[-\pi \frac{C_3^2}{\Omega\sqrt{C_2^2-C_1^2}}]=exp[-2\pi\delta]$, say.
Thus $p_z=0$ implies $P_{LZ}=1$ and $C_3\rightarrow\infty$ implies $P_{LZ}=0$.
Away from $t_1~\&~t_2,$ the spectrum enters into the impulse regime where transition in not allowed for small $\Omega$. The averaged transition probability for sweeping through a full cycle becomes $\bar{P_{(1)}}=2P_{LZ}(1-P_{LZ})$, for there are two ACPs per cycle\cite{lz,kar1,kar2}. One also need to consider the contribution from the Stuckelberg interference\cite{lz} between the two probabilistic pathways. With this, the actual excitation probability becomes $P_{(1)}=4P_{LZ}(1-P_{LZ})sin^2\Phi_{st}$ with $\Phi_{st}=\xi_2+\phi_S$ being the Stuckelberg phase where $\xi_2=\int_{t_2}^{t_1+2\pi/\Omega}\epsilon_{n,p_z}dt$ and $\phi_S=\delta(ln\delta-1)+arg\Gamma(1-i\delta)-\pi/4$\cite{kar2}. So both $P_{LZ}=0$ and $P_{LZ}=1$ cases lead to zero transition with $P_{(1)}=0$. Rather for fractional $P_{LZ}$ values, one can observe finite transition probabilities across the full cycle. 
We should add here that multiple passage through the ACPs in a stroboscopic fashion gives an overall transition probability\cite{kar2} (for $m$ cycles) to be $P_{(m)}=P_{(1)}\frac{sin^2(m\phi)}{sin^2\phi}$ where $cos\phi=-(1-p)cos(\xi_1+\xi_2+2\theta)-p cos(\xi_1-\xi_2)$.
\begin{figure}
\centering
\includegraphics[width=\linewidth,height=2.5 in]{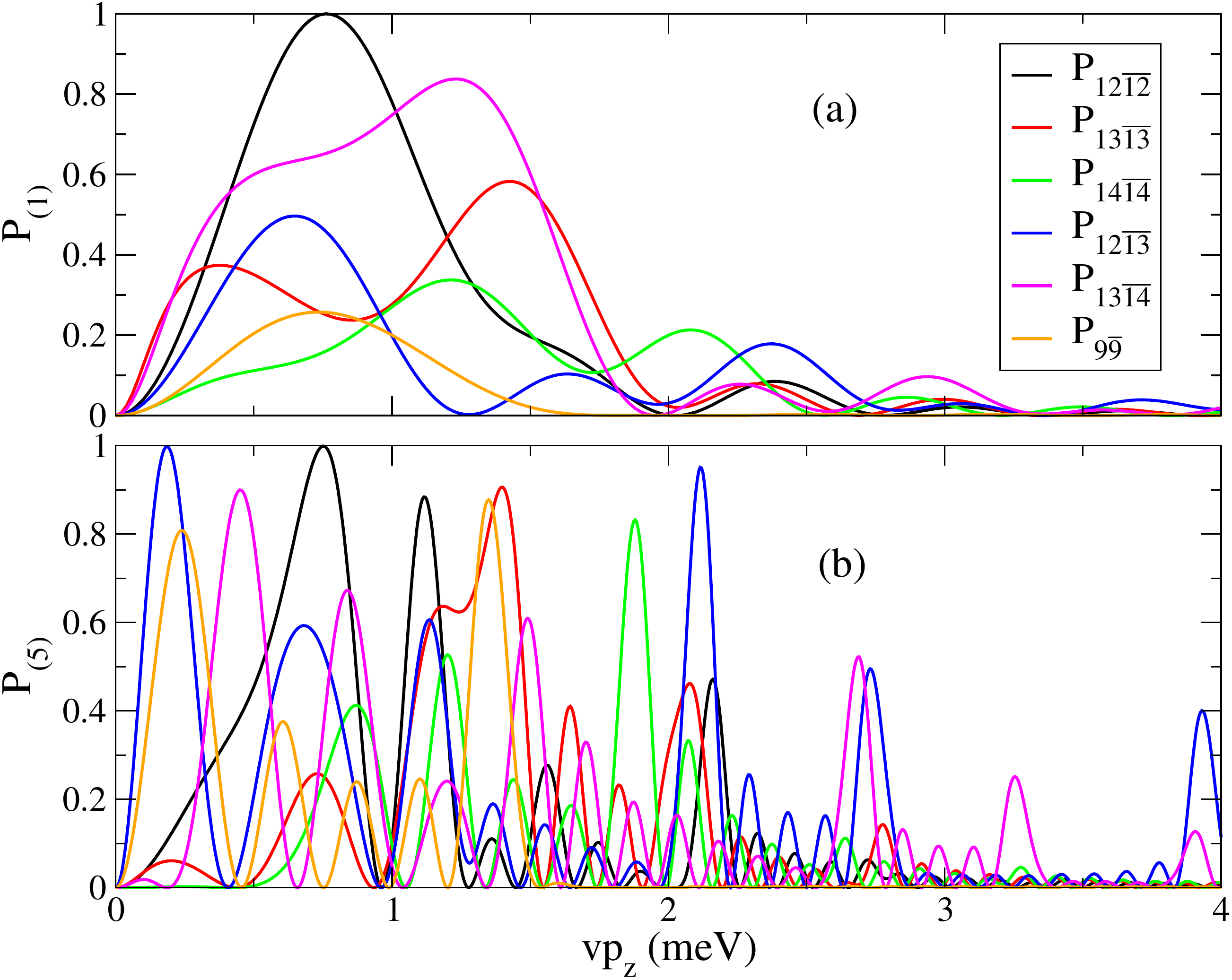}\\
\includegraphics[width=\linewidth,height=1.4 in]{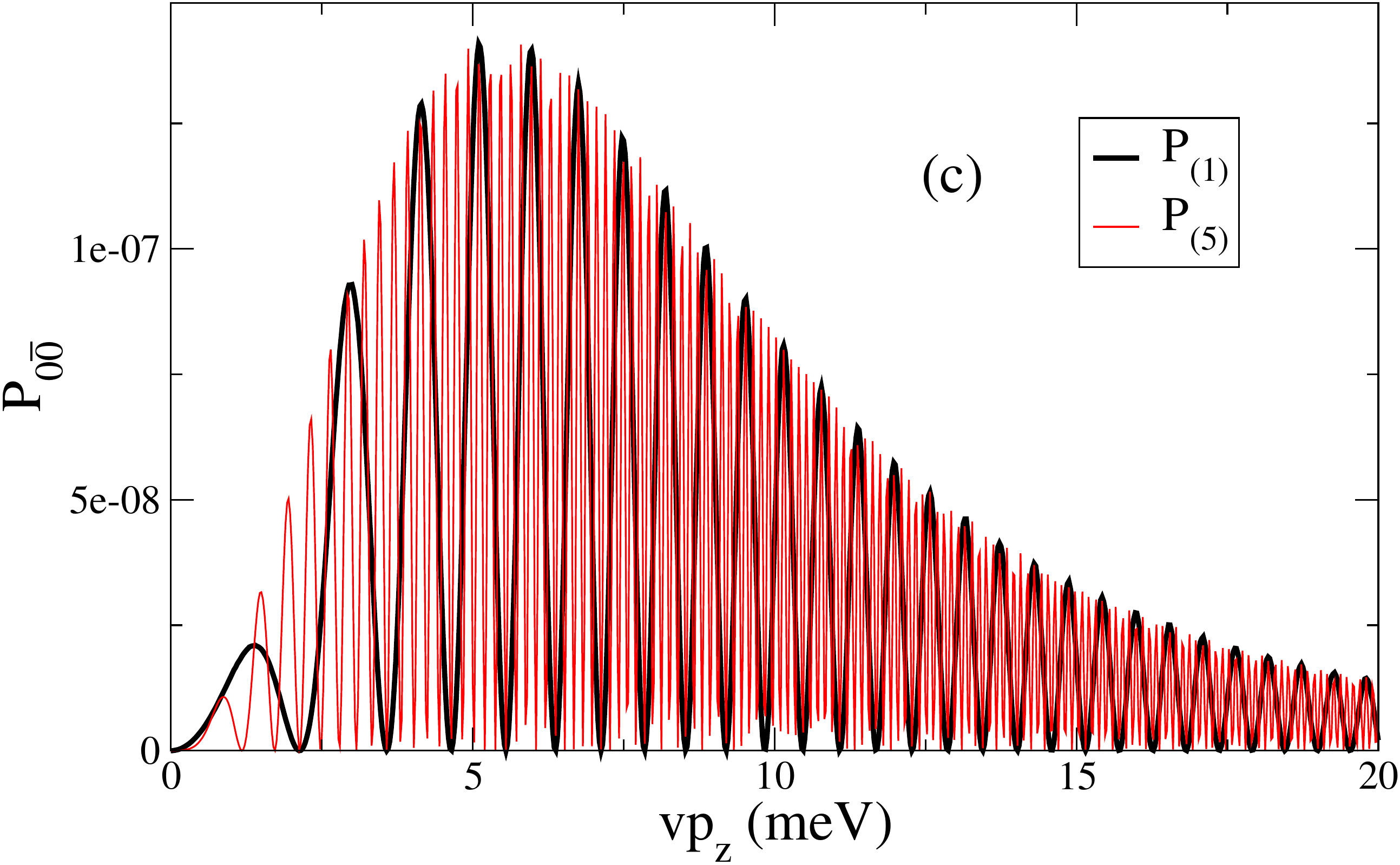}
\caption{(Color online) The probability of intra/inter band transitions between low energy levels for $B_0=5~T,~B_1=2~T$ and $m_0=8~meV$. {(a) and (b) show transition probabilities $P_{12,\bar{12}},~P_{13,\bar{13}},~P_{14,\bar{14}},~P_{12,\bar{13}},~P_{13,\bar{14}},~P_{9,\bar 9}$ as a function of $vp_z$ for one ($P_{(1)}$) and five ($P_{(5)}$) sweeps of field respectively. (c) gives $P_{0,\bar 0}$ variations for one and five sweeps of the field.}}
\label{figlz}
\end{figure}

Adiabatic-Impulse picture is not tenable in large $\Omega$ limit. {Though here in this paper we consider only small $\Omega$ values, for the continuity of discussion we briefly mention that} one can also analytically obtain the transition probabilities {for large $\Omega$} using a rotating wave approximation that utilizes suitable rotation of the basis states of the TLS Hamiltonian\cite{lz,kar2}.
We can also resort to the Floquet formalism\cite{floquet} to get the Floquet spectra of the problem that gives an effective static Hamiltonian
\begin{eqnarray}
  H_{eff}=C_1\sigma_z+C_3\sigma_y+\frac{1}{\hbar\Omega}2iC_2C_3\sigma_x
\end{eqnarray}
for the stroboscopic dynamics of the problem.
This indicates impossibility of gaplessness of the Floquet spectrum at $p_z=0$ as $C_1\neq0$. The Stuckelberg phase effects are averaged out due to fast driving of the ac field. One can still obtain finite transitions between the Floquet modes.

{Fig.\ref{figlz}(a)-(b) display typical excitation probabilities due to one and five sweeps of the field obtained numerically for transition from VB to CB at low lying Landau states. See that all states are gapped for $p_z\ne0$ whereas for $p_z=0$, LL states can be both gapped ($eg.,~n=9$) and gapless ($eg.,~n=12,13$ and $14$). As discussed before, small gaps pave for electronic transitions between the levels whereas LL states that become gapless twice within every cycle of field oscillation lead to zero transition probabilities. The excitations amounting to level crossing in a TLS are called defect productions\cite{lz,kar1,kar2} in nonequilibrium dynamics problems and on average, the time periodic spectral profile allows quantum tunneling of finite fraction of VB electrons to CB altering the system's conductance\cite{lzcond,lzcond2,lzcond3}.} We should add here that transitions need not necessarily be of intraband type as electrons can transit between different LL states as well. {In Fig.\ref{figlz}(a)-(b), we also show interband transition probabilities $P_{n{\bar l}}$ for transitions from the $\bar l$ (a VB state) to the $n$ (a CB state) LL state with $(n,l)$=(12,13) and (13,14). Now these results can be contrasted with  LZ transition probabilities between n=0 VB and CB LL states in Fig.\ref{figlz}(c). This shows almost zero transition probabilities due to large gap between the zero LL states.} 

\section{Field parallel to the nodal plane}
{Next when we consider the magnetic field to be parallel to the nodal loop, spectral characteristics change drastically. Apart from trivial magnetic oscillations, here we can also witness topologically nontrivial oscillations corresponding to $\gamma=0$ and Berry phase equal to $\pm\pi$\cite{cortijo}.}

With $B$ along the $x$ direction, cyclotron orbits appear in the $y-z$ planes. Choosing ${\bf A}=(0,-Bz,0)$, the Hamiltonian becomes
\begin{eqnarray}
H(B,p_x)=(\frac{p_x^2+(p_y+eBz)^2}{2m}-m_0)\sigma_z+vp_z\sigma_y
\label{eq1b}
\end{eqnarray}
This can not be diagonalized analytically but dispersions can be obtained semi-classically that fits well with numerical solutions\cite{cortijo}.
{We consider extremal orbits (corresponding to dispersion extrema or $\partial A_n/\partial p_x=0$) that appear at $p_x=0$.
  These are obtained from the intersections of the $p_x=0$ plane and the Fermi surface.}

{To capture the QO characteristics, we particularly study the 2D Hamiltonian $H(B,p_x=0)$.
To begin with the field-free case (i.e., {\bf B=0}) within the $p_x=0$ plane,
 we find that the band touching points appear at $p_z=p_{z0}=0,p_y^2=p_{y0}^2=2mm_0$. Linearizing about those nodes one can get the low energy Hamiltonian $H=\frac{p_{y0}}{m}p_y'\sigma_z+vp_z'\sigma_y$ where $p_{y(z)}'=p_{y(z)}-p_{y0(z0)}$ are the reduced variables. This represents an anisotropic Dirac Hamiltonian about the pair of Dirac points $(0,\pm\sqrt{2mm_0},0)$ (and they merge into a semi-Dirac point for $m_0=0$)\cite{cortijo}.}

\begin{figure}
\centering
\includegraphics[width=.32\linewidth]{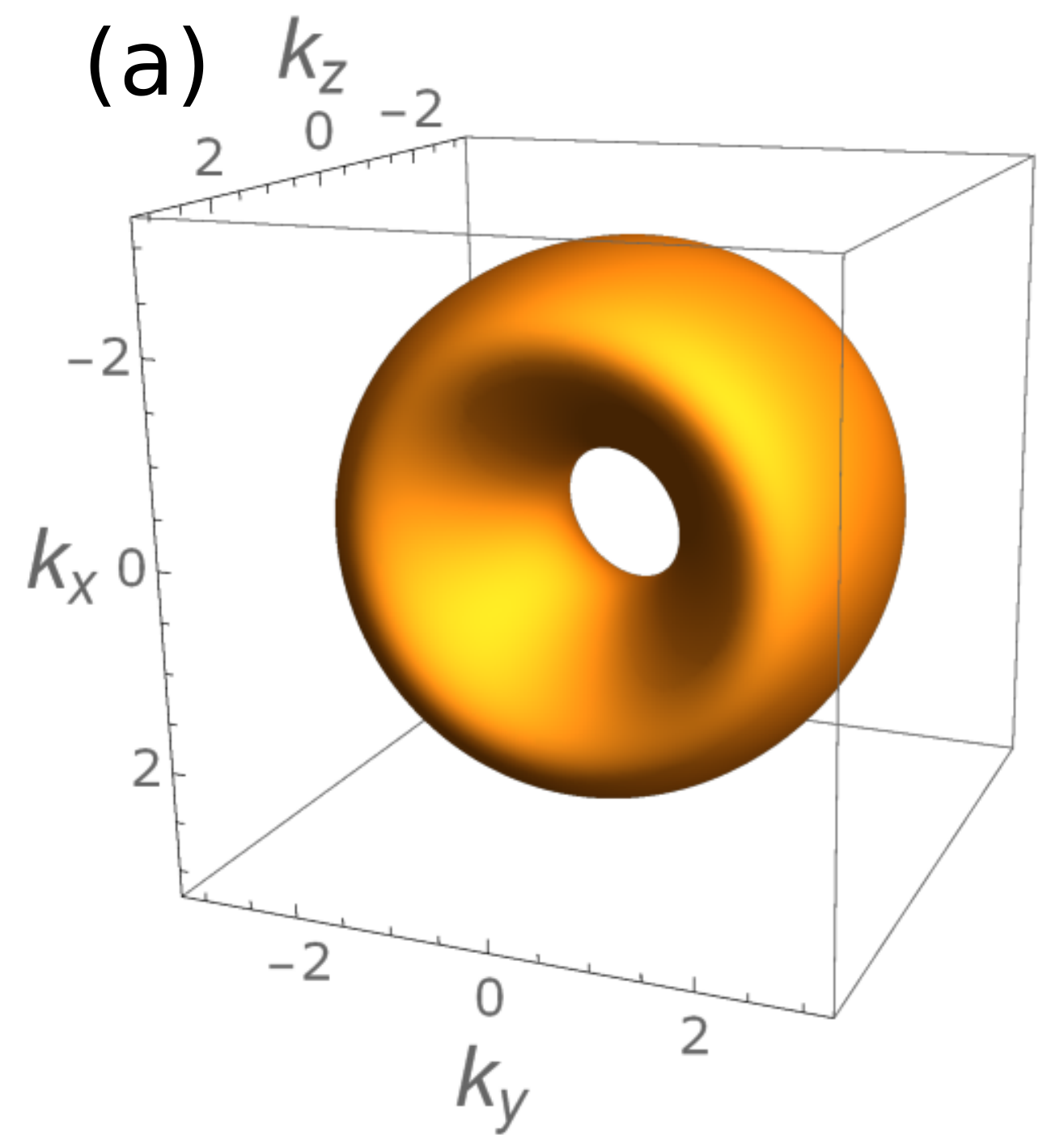}\hskip .05 in \includegraphics[width=.32\linewidth]{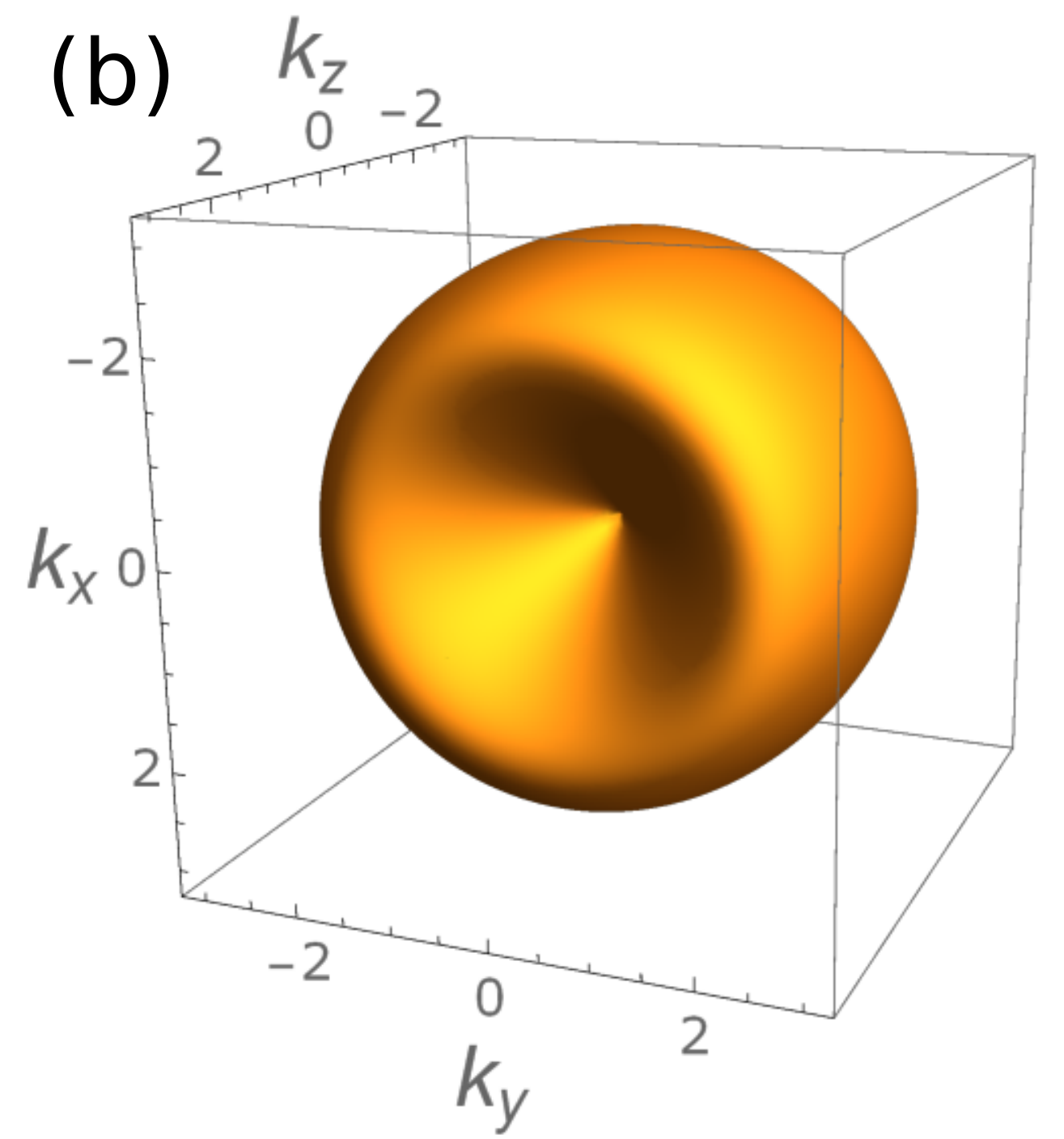}\hskip .05 in\includegraphics[width=.32\linewidth]{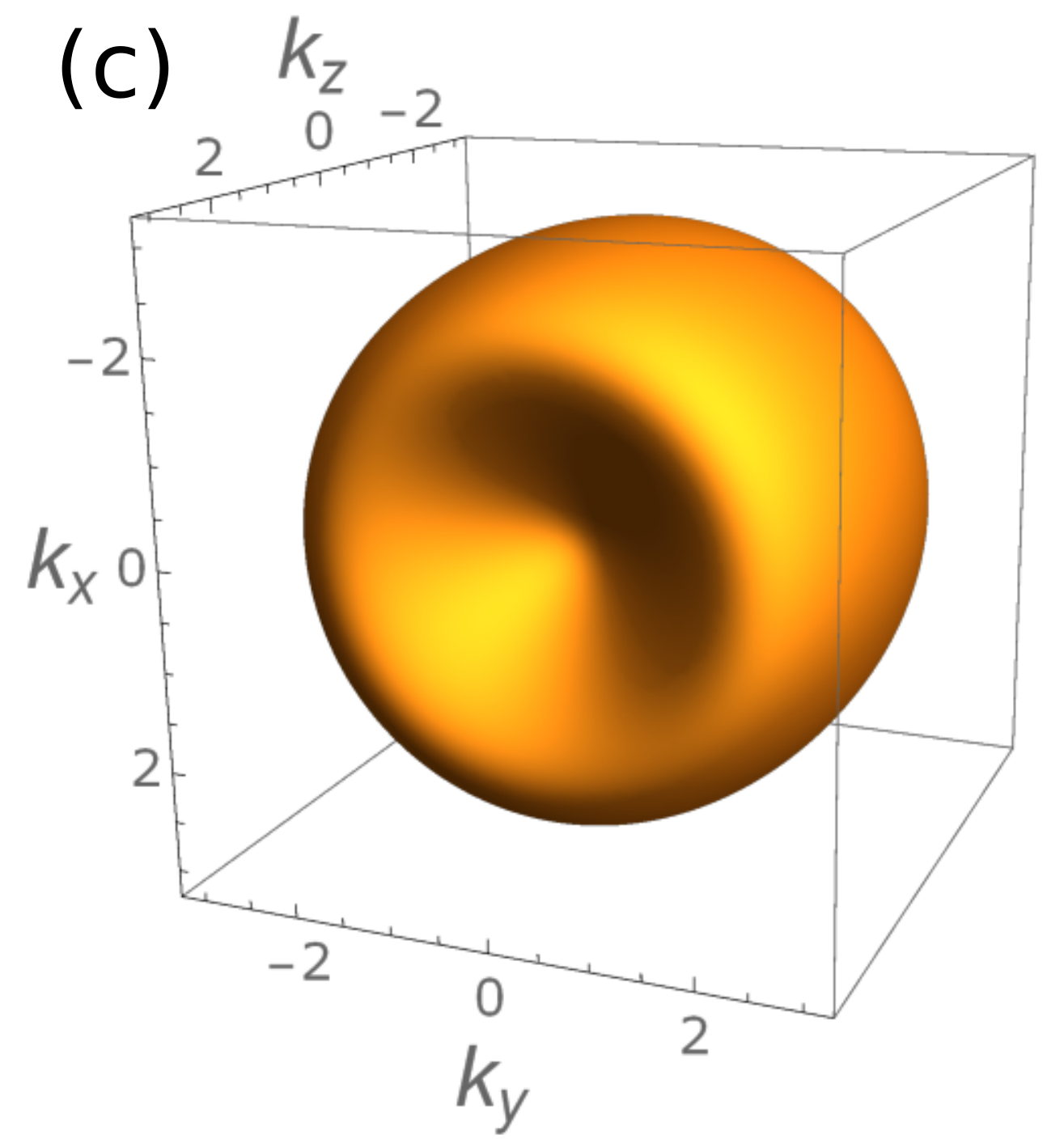}\\\vskip .1 in
\includegraphics[width=.32\linewidth]{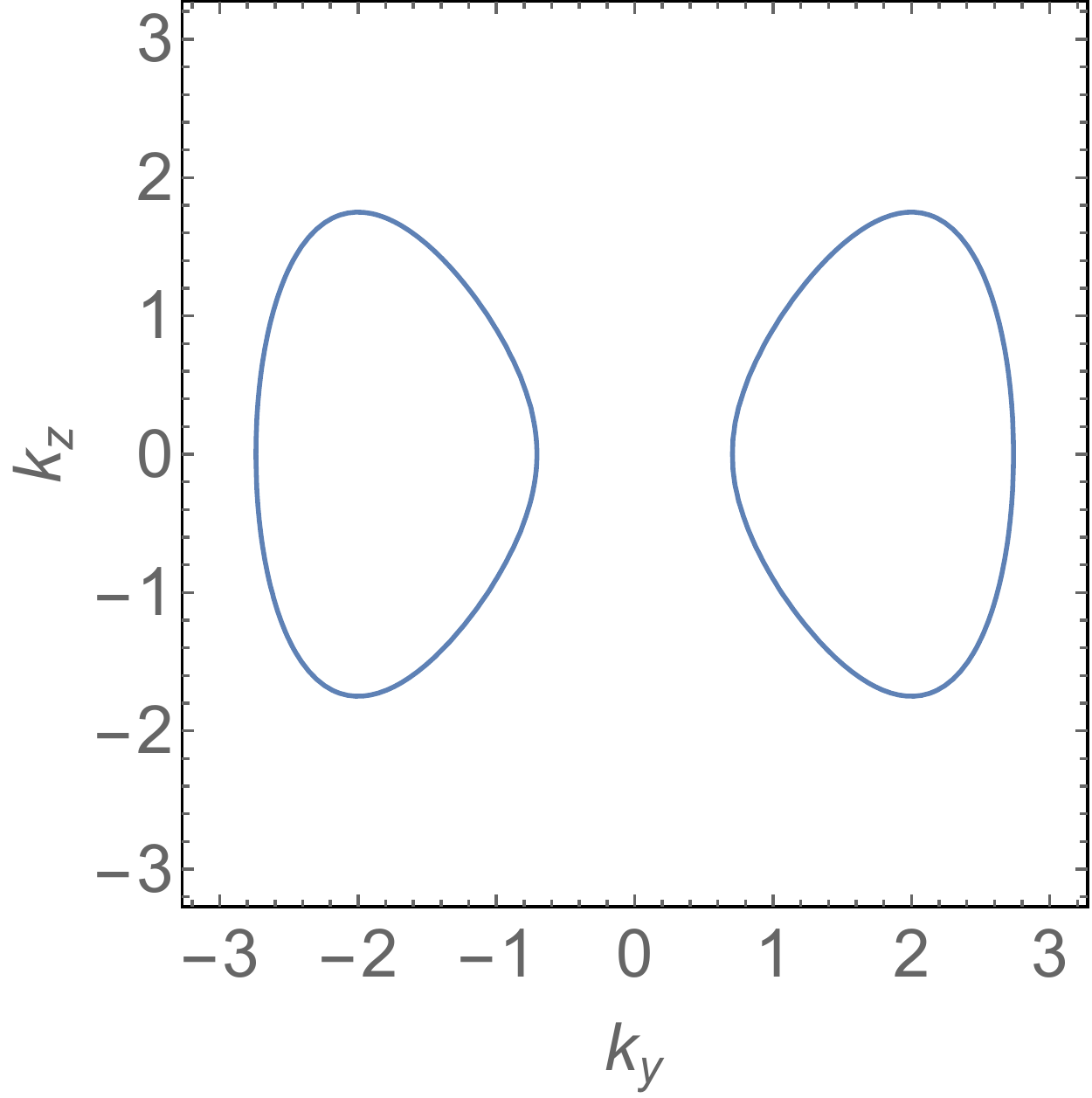}\hskip .05 in\includegraphics[width=.32\linewidth]{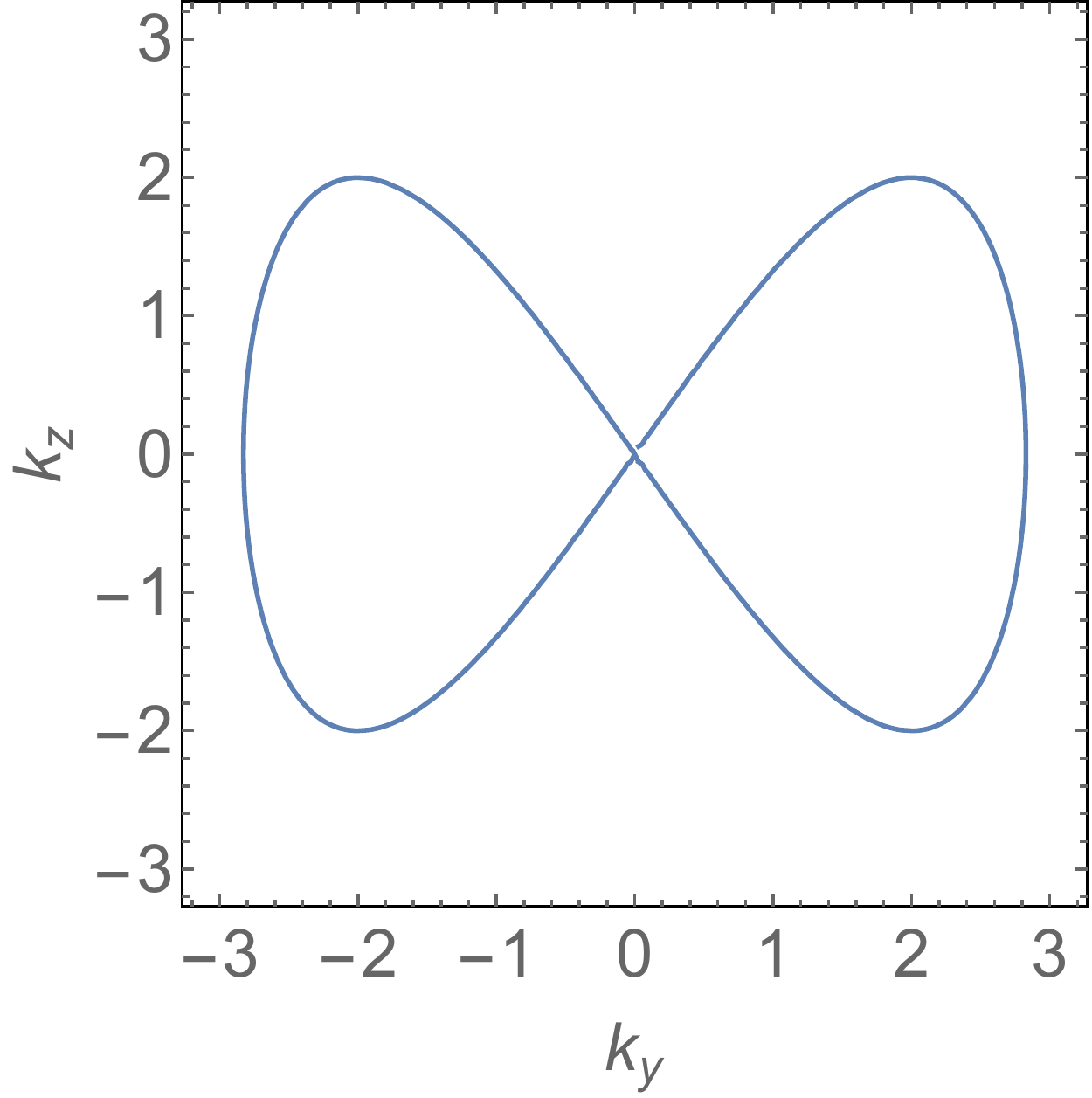}\hskip .05 in\includegraphics[width=.32\linewidth]{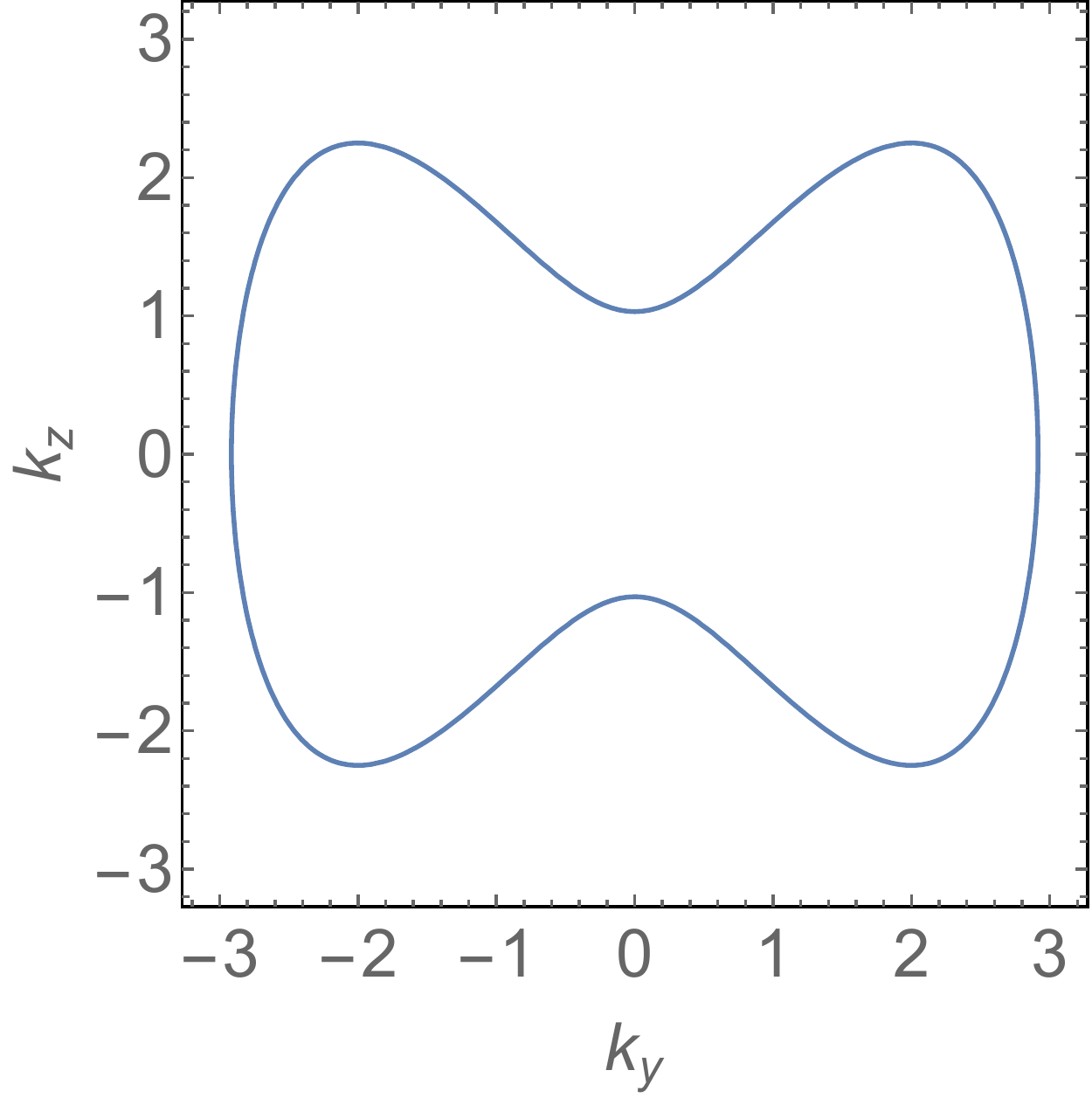}
\caption{{(Color online) Top panel shows typical Fermi surfaces (FS) for (a) $\mu<m_0$, (b) $\mu=m_0$ and (c) $\mu>m_0$ (here $k_{x(y,z)}=p_{x(y,z)}/\hbar$). Notice that FS changes topology at $\mu=m_0$. The bottom panel shows respective semiclassical extremal orbits at $p_x=0$ when ${\bf B}=B{\bf \hat x}$.}}
\label{taurus}
\end{figure}

In presence of ${\bf B}=B{\hat x}$, new terms due to Peierls substitution are added to the Hamiltonian.
With substitution $\tilde{z}=z+\frac{p_y}{eB}$, the Hamiltonian can be rewritten as
\begin{align}
  H(B,p_x=0)&=(\frac{m\omega_c^2{\tilde{z}}^2}{2}-m_0)\sigma_z+vp_z\sigma_y\nonumber\\&=[\frac{m\hbar^2\omega_c^2v^2}{2}]^{1/3}[(Z^2-\delta)\sigma_z+P\sigma_y]
  \label{eqB||}
\end{align}
where $\delta=[\frac{2m_0^3}{m\hbar^2\omega_c^2v^2}]^{1/3}$, $Z=\frac{\tilde z}{\alpha\sqrt\hbar}$ and $P=\frac{\alpha p_z}{\sqrt\hbar}$ are dimensionless parameters with $\alpha=[\frac{2v}{m\omega_c^2\sqrt\hbar}]^{1/3}$.
{This shows the dispersion} $\epsilon_n(B,p_x=0)=[\frac{m\hbar^2\omega_c^2v^2}{2}]^{1/3}\sqrt{E_n(B)}$, $E_n(B)$ being the eigenvalues of an anharmonic oscillator Hamiltonian {(see {appendix-B}). We find}
$\epsilon_n(B,0)\simeq\pm 2[\frac{m_0}{2m}]^{1/4}\sqrt{veB\hbar n}$ at low $B$ and low energies\cite{B@x}. {It indicates doubly degeneracy not only in gapless $n=0$ mode but also in gapped $n\ne0$ modes.
  In this regime, the LL states show valley degeneracy corresponding to the pair of valleys with opposite chiralities at the two Dirac points and a semiclassical orbit about a Dirac point picks up a Berry phase $\gamma_B=\pm\pi$ indicating a topological order\cite{lim2} (see {appendix-B}).
  However for large energies, the approximation used in getting the expression $\epsilon_n(B,0)$ wears off breaking the degeneracy of the $n\ne0$ modes of the system.
Particularly for $\mu\gtrsim m_0$ ($m_0$ being of the order of the intervalley barrier), the semi-classical orbit encloses both the Dirac points yielding an overall zero Berry phase\cite{B@x}.
One can find that the ring torus of Fermi surface to change into spindle torus for large chemical potential values\cite{B@x}.} {Fig.\ref{taurus} gives a pictorial description of the Fermi surfaces of the present system for $\mu<,=,>m_0$ to show how it changes topology at $\mu=m_0$. In all three scenerio, magnetic oscillations remain topologically trivial for ${\bf B}=B{\bf \hat z}$ (as the cyclotron orbits do not cut through the nodal ring). But for ${\bf B}=B{\bf \hat x}$, topological oscillations are observed at low energies. There the extremal semiclassical orbits at $p_x=0$ become disjoint loops about the two Dirac points. But those orbits merge into a single extended loop for $\mu>m_0$ (this, however, does not imply the Dirac points to merge with each other, which would happen only if $\delta\rightarrow 0$) turning the QO phenomena topologically trivial.}

{One can also find the signature of such topological transitions also from the DOS which shows divergence (a Van Hove singularity) at the transition point $\mu\sim m_0$\cite{B@x}.} For a $\mu$ below such critical value, one get two extremal orbits of same area in the Landau fan diagram and each of them ring through the nodal loop resulting in topological $\pm\pi$ Berry phases. Above the critical $\mu$, the orbit encircles both the gapless points thereby giving zero winding or Berry phase.
\begin{figure}
\centering
\includegraphics[width=\linewidth,height=2.5 in]{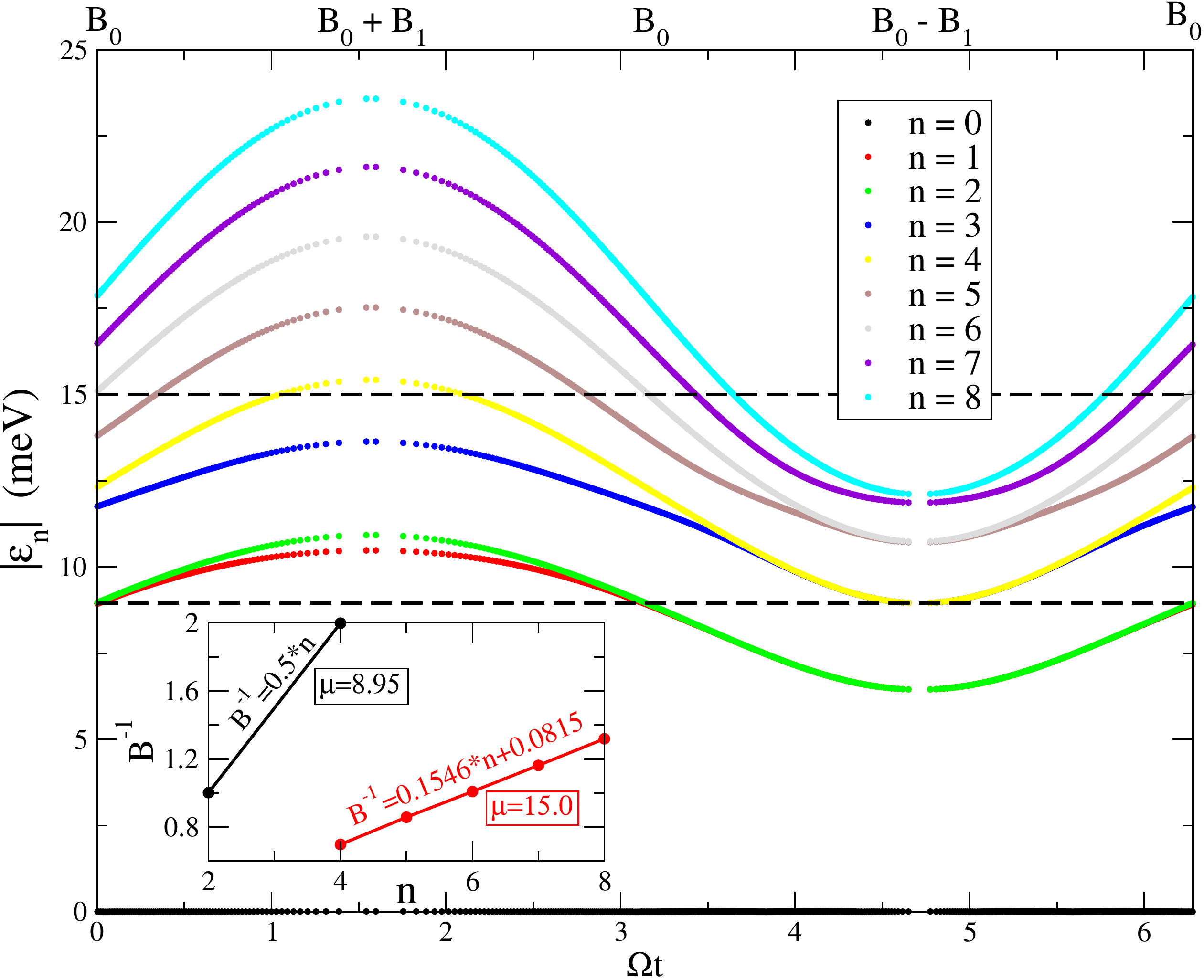}
\caption{(Color online) Plot of (top) the energy dispersion for low energy Landau states for $p_x=0$ for parameters $B_0=1~T,~B_1=0.5~T$ and $m_0=12$ meV. {Few field values at different times are mentioned at the top. Notice the intersections of the $\mu=8.95$ and $15.0~meV$ lines (dashed) with the dispersion spectra.} The inset shows Landau fan diagrams for those two $\mu$ values.}
\label{lfan2}
\end{figure}

When we consider the oscillating field $B(t)=B_0+B_1sin(\Omega t)$, {a few of the Landau level dispersions can be made to periodically pass through both the topologically different regimes of $\epsilon_n>m_0$ and $\epsilon_n<m_0$} if suitable values of $B_0$ and $B_1$ are chosen.
{Fig.\ref{lfan2}} shows the numerical spectra for low lying LL states as a function of time for one sweep of the ac field. With the choice of $B_0=1~T,~B_1=0.5~T$ and $m_0=12~meV$, here we witness both high energy topologically trivial and low energy topologically non-trivial magnetic oscillations and their transitions between each other as the field is swept through.
{One can see the degeneracy breaking of the $n\ne0$ modes at high energies $\epsilon_n\gtrsim m_0$. A periodic variation in field, that causes a periodic variation in dispersions, causes the system to switch between regimes with degenerate and nondegenerate LL spectra and hence between topological and trivial oscillations.} {Fig.\ref{lfan2} also shows two equi-energy lines of $\mu=8.95$ and $15.0~meV$ (below and above $\mu=m_0$ respectively) which can intersect only a limited number of LL spectra (depending on the field parameters chosen).} The inset shows the $B^{-1}$ versus $n$ Landau fan diagrams. {There $n$-intercepts$\sim$ 0.0 and -0.5 are obtained for topologically trivial low energy ($\mu=8.95~meV$) and topologically nontrivial high energy ($\mu=15.0~meV$) extremal orbits respectively. A comparison with the Onsager relation indicate $\gamma=0$ and $0.5$ for these two respective cases. Also the slopes indicate larger QO frequencies for low energy topological oscillations when compared with the high energy trivial oscillations.} One also should notice that, unlike the previous case with $B\perp$ nodal-plane, here $n=0$ state remains the lowest energy state which is gapless for small $B$ values.


\section{Conclusion}
{The present work proposes a time periodic quench to a NLSM system realized via periodic strong magnetic field that causes quantum oscillation to be observed in a controlled manner and with typicalities of sinusoidal variation of the field.} In this paper we analyze both spectral and topological response of the field acting on a NLSM. Firstly for the field perpendicular to the nodal ring, we find the quantized Landau level states to periodically alter their dispersions with time and the temporal periodicity of the quantum oscillation gets modified accordingly. {For low lying modes, repeated proximity of the VB and CB causes inter/intraband transitions to take place that gives rise to new routes to electron conduction.} {Trivial magnetic oscillations are registered for $B\perp$ nodal-plane} with zero Berry phases for the semiclassical cyclotron orbits. But for a magnetic field in the nodal plane,
we obtain topologically nontrivial QO at low energies, though it changes back to trivial oscillation at large energies. {With carefully chosen field parameters one can thus allow the Landau level spectra to transit between trivial and non-trivial phases periodically} and leave room for many exotic spectral/transport phenomena to explore.
{For example, NLSM compound $ZrSiS$ or $CaAl_4$ show strong de-Haas van Alphen (dHvA) oscillations at low temperatures, visible clearly after removing the paramagnetic background. Ref.\onlinecite{expt,exptb} report dHvA oscillations in magnetization in these compounds for field upto 7 T or 14 T respectively with both $B||ab$ and $B\perp ab$ plane.
It will be interesting to study the effect of oscillating magnetic field acting on such systems and the evolution of the nontrivial Berry phases that become topological for $\gamma_B=\pm\pi$.
There are also scopes for further work if we set $B_0=0$} and thus allow for low $B$ values to see how Zeeman effects alter the spectral and QO results\cite{exptc} at low energies.
In a later communication, we also plan to study {in detail how the inter/intra band transitions affect the conductivity and hence the electronics of the system.}

\section*{Acknowledgements}
{SK thanks M. Goerbig, S. Mandal, D. Sinha, B. Basu and A. Jayannavar for useful discussions and feedbacks.} {SK also acknowledges financial support from DST-SERB, Government of India under grant no. SRG/2019/002143.}

{
  \section*{Appendix-A}
For an electron in an electromagnetic field, one can construct the Lagrangian and the Hamiltonian starting from the Lorentz force equations\cite{cohen}. The magnetic field enters the Hamiltonian only via canonical momenta and accordingly $H$ gets a substituted form:
  \begin{eqnarray}
    p^2/2m\rightarrow(p-eA)^2/2m+e\phi
    \label{peierl}
  \end{eqnarray}
  in presence of the electromagnetic field (or $p^2/2m\rightarrow(p-eA)^2/2m$ for only a magnetic field) where $\phi$ denotes the scalar potential. This is true both in classical and quantum systems\cite{cohen}. Also see that it holds for both time independent as well as time dependent vector potentials. As the present problem considers a magnetic field alone, we can choose $\phi$ to be zero and the Hamiltonian, in presence of the magnetic field, takes the form as in Eq.\ref{eqB}.}
  
 {Now notice that the Eq.\ref{eqB} can be rewritten as
 \begin{eqnarray}
   H&=&[\frac{p_y^2}{2m}+\frac{m}{2}(\frac{eB}{m})^2(y-y_0)^2-m_0]\sigma_z+vp_z\sigma_y
   \label{lho}
 \end{eqnarray}
 Here the first two terms (apart from the pseudospin $\sigma_z$) represent a 1D Hamiltonian for displaced Harmonic oscillator along $y$ direction with centers at $y_0={p_x/eB}$. The corresponding eigenfunctions constitute the Landau basis given by\cite{fazekas} 
 $\psi^L_n\sim e^{ip_xx/\hbar}H_n(\frac{y-y_0}{l})e^{-(y-y_0)^2/2l^2}$ where $H_n$ denotes the Hermite polynomial of $n$-th order and $l=\sqrt{\hbar/eB}$ is the magnetic length. Following operator algebra\cite{cohen}, one can write this linear Harmonic oscillator Hamiltonian as $({\hat n}+1/2)\hbar\frac{eB}{m}$ that satisfies ${\hat n}\psi^L_n=n\psi^L_n$. Similarly the full Hamiltonian matrix in Eq.\ref{lho} can be expressed in a basis given by $\psi_n^\pm\sim\psi^L_ne^{ip_zz/\hbar}\hat{\eta}^\pm$ where $\sigma_z\hat{\eta}^\pm=\pm\hat{\eta}^\pm$. This let the Hamiltonian \ref{lho} to transform into Eq.\ref{eqLL} (with the $\psi^L_n$ factor of the basis function transforming the Harmonic oscillator part of the Hamiltonian to $({ n}+1/2)\hbar\frac{eB}{m}$). The diagonalization of the resulting matrix gives the energy eigenvalues as $\epsilon=\pm\sqrt{[(n+\frac{1}{2})\frac{e B\hbar}{m}-m_0]^2+v^2p_z^2}$ (see also Ref.\onlinecite{molina}).}

 {We should mention here that a time dependent ${\bf B}$ also produces an electric field ${\bf E}=-\partial{\bf A}/\partial t-\nabla\phi$.
 With $\phi=0$, no changes appear in the Hamiltonian. But the time dependence results a non-zero electric field ${\bf E}=-\partial{\bf A}/\partial t=yB_1\Omega~cos(\Omega t)$ that contributes to the conduction. If we want to ignore that, $\Omega$ has to be small. A comparison with the magnetic field reveals that a small ${\bf E}$ implies ${yB_1\Omega cos(\Omega t)<<vB}$ or more strictly $\Omega<<\frac{vB}{y_{max}B_1}$, $y_{max}$ denoting the maximum value of $y$ in the NLSM sample considered.}
{\section*{Appendix-B}
On squaring Eq.\ref{eqB||}, one obtains (after an unitary transformation) an anharmonic oscillator Hamiltonian
\begin{align}
  H^2(B,0)/[\frac{m\hbar^2\omega_c^2v^2}{2}]^{2/3}&=((Z^2-\delta)^2+P^2-i[P,Z^2])\sigma_0\nonumber\\=&((Z^2-\delta)^2+P^2-2Z)\sigma_0.
  \label{}
\end{align}
For large $\delta$ ($i.e.,$ small $B$ values), the spectral minima correspond to $Z=\pm\sqrt{\delta}$ about which the low energy eigenvalues can be obtained as $E_n(\delta)=4n\sqrt{\delta}$\cite{B@x}.
So the discrete spectra contains a gapless $n=0$ mode and doubly degenerate gapped $n\ne0$ modes.
The degeneracy, however, breaks as $\delta$ becomes small when the two energy minima come close enough to be treated independently.
This is like a double well potential problem with well separated pair of minima.
However, a large $B$ (or small $m_0$) bring the minima closer as well as reduces the intermediate potential barrier. This enables quantum tunneling between the valleys there by breaking the valley degeneracy\cite{cortijo}. {Furthermore, it also opens a gap in the $n=0$ mode.}
 From the Berry phase calculations one can show that such degeneracy breaking brings in a topological change in the system.}

{For $B=B\hat z$, Landau levels correspond to semiclassical circular orbits normal to the $\hat z$ direction. The Bloch functions\cite{ashcroft} or two-component spinors at $p_z=0$ plane are independent of momentum $p$ which can be chosen as $u_p=(1,0)^T$. Hence they produce no winding in the spinor structure\cite{cortijo} or Berry phase for the closed electronic orbits. However for $B=B\hat x$, $u_p$ has a $p$ dependence: $u_p=\frac{1}{\sqrt{2}}(1,e^{i\phi_p})^T$, with
  $\phi_p=tan^{-1}(mvp_z'/p_{y0}p_y')$ around the Dirac points. This causes $\pm\pi$ Berry phases obtained as closed line integral of the Berry connection $<u_p|i\nabla_p|u_p>$ along the extremal cyclotron orbits (which is now elliptical due to the anisotropy) around individual Dirac points. However for smaller $\delta$ values, the Dirac point pairs come closer to merge finally into a semi-Dirac point at $\delta=0$ where Berry phase become zero. Even for small nonzero $\delta$, one can consider high energies for which the electron orbits circle around both the Dirac points providing an overal zero Berry phase to the system\cite{B@x}.
}

\end{document}